**Bogoliubov Quasiparticle on the Gossamer Fermi Surface in Electron-Doped Cuprates**

**Authors:**

Ke-Jun Xu[1,2,3]*, Qinda Guo[4]*, Makoto Hashimoto[5]*, Zi-Xiang Li[6,7]*, Su-Di Chen[1,2,3]†, Junfeng He[1,2,8], Yu He[1,2,3,9], Cong Li[3,4], Magnus H. Berntsen[4], Costel R. Rotundu[1,2], Young S. Lee[1,2,3], Thomas P. Devereaux[1,2,10], Andreas Rydh[11], Dong-Hui Lu[5], Dung-Hai Lee[12,13]‡, Oscar Tjernberg[4]‡, and Zhi-Xun Shen[1,2,3,14]‡

[1] Stanford Institute for Materials and Energy Sciences, SLAC National Accelerator Laboratory, 2575 Sand Hill Road, Menlo Park, California 94025, USA

[2] Geballe Laboratory for Advanced Materials, Stanford University, Stanford, California 94305, USA

[3] Department of Applied Physics, Stanford University, Stanford, California 94305, USA

[4] Department of Applied Physics, KTH Royal Institute of Technology, Hannes Alfvéns väg 12, 114 19 Stockholm, Sweden

[5] Stanford Synchrotron Radiation Lightsource, SLAC National Accelerator Laboratory, 2575 Sand Hill Road, Menlo Park, California 94025, USA

[6] Beijing National Laboratory for Condensed Matter Physics and Institute of Physics, Chinese Academy of Sciences, Beijing 100190, China

[7] University of Chinese Academy of Sciences, Beijing 100049, China

[8] Department of Physics and CAS Key Laboratory of Strongly-coupled Quantum Matter Physics, University of Science and Technology of China, Hefei, Anhui 230026, China

[9] Department of Applied Physics, Yale University, New Haven, Connecticut 06511, USA

[10] Department of Materials Science and Engineering, Stanford University, Stanford, California 94305, USA

[11] Department of Physics, Stockholm University, SE-106 91 Stockholm, Sweden

[12] Department of Physics, University of California, Berkeley, California 94720, USA

[13] Material Sciences Division, Lawrence Berkeley National Laboratory, Berkeley, California 94720, USA

[14] Department of Physics, Stanford University, Stanford, California 94305, USA

* These authors contributed equally to this work.

† Current address: Kavli Energy NanoScience Institute, University of California, Berkeley, CA 94720, USA

‡ Corresponding authors: D.-H. Lee (dunghai@berkeley.edu), O. T. (oscar@kth.se), Z.-X. S. (zxshen@stanford.edu)

**In contrast to hole-doped cuprates, electron-doped cuprates consistently exhibit strong antiferromagnetic correlations with a commensurate ($\pi$, $\pi$) ordering wave vector[1,2], leading to the prevalent belief that antiferromagnetic spin fluctuations mediate Cooper pairing in these unconventional superconductors [3]. However, early investigations produced two paradoxical findings: while antiferromagnetic spin fluctuations create the largest pseudogap at "hot spots" in momentum space [4,5,6,7,8,9,10], Raman scattering [11] and angle-resolved photoemission spectroscopy measurements using the leading-edge method[12] seem to suggest the superconducting gap is also maximized at these locations. This presented a dilemma for spin-fluctuation-mediated pairing: Cooper pairing is strongest at momenta where normal state low energy spectral weight is most suppressed. Here we investigate this dilemma in $Nd_{2-x}Ce_xCuO_4$ using angle-resolved photoemission spectroscopy under significantly improved experimental conditions. The unprecedented signal-to-noise ratio and resolution allow us to directly observe the Bogoliubov quasiparticles, demonstrating the existence and importance of two sectors of states: (1) The reconstructed main band and the states gapped by the antiferromagnetic pseudogap around the hot spots. (2) The "gossamer" Fermi surface states with distinct dispersion inside the pseudogap, from which Bogoliubov quasiparticle coherence peaks emerge below $T_c$. Supported by numerical results, we propose that the non-zero modulus of the antiferromagnetic order parameter causes the former, while fluctuations in the antiferromagnetic order parameter orientation are responsible for the latter. Moreover, the largest superconducting gap, derived from Bogoliubov quasiparticle energy, is found to be an order of magnitude smaller than the pseudogap, establishing the distinct nature of these two gaps. Our revelations of the gossamer Fermi surface reconcile the paradoxical observations, deepening our understanding of superconductivity in electron-doped cuprates in particular, and unconventional superconductivity in general.**

**Main**

Bardeen-Cooper-Schrieffer's (BCS) theory of superconductivity [13] successfully explains the pairing and condensation of quasiparticles in conventional superconductors, where the attractive potential that binds the electrons into Cooper pairs arises from the exchange of lattice vibrations or phonons. However, in unconventional superconductors like the cuprates, the pairing "glue" is thought to arise from the electrons themselves through electron-electron interactions[14,15,16,17]. This means that the electrons simultaneously act "to glue and to be glued". An early discussion on the challenge of the dual roles was given in Ref. 18.

Because the parent compound of the cuprates are antiferromagnetic (AF) insulators and superconductivity arises after the demise of AF long-range order[3], from early on it has been proposed that AF spin fluctuations are important source of Cooper pairing in cuprate superconductors. In addition to the cuprates, other materials where AF is suspected to play an important role in pairing include iron-based superconductors, heavy-fermion superconductors, organic superconductors, etc[15,16,17]. Therefore, acquiring a comprehensive understanding of the mechanism behind AF spin fluctuations is essential in unconventional superconductivity research.

For hole doped cuprates, spin fluctuation exchange is predicted to cause the d-wave pairing symmetry[14], which is confirmed by experiments [19,20]. However, due to the presence of pseudogap [21,22,23,24], multiple competing ordering tendencies [25,26,27,28,29,30,31], Van Hove singularity[32], and phenomena associated with antinodal $B_{1g}$ phonon coupling[33,34,35], unambiguous identification of the pairing mechanism remains a daunting task. The situation is significantly different in the electron-doped cuprates - the AF correlation is much stronger than any other ordering tendencies[3]. This makes them excellent model systems to study the spin fluctuation pairing mechanism.

**$Nd_{1.85}Ce_{0.15}CuO_4$ as a model system**

The magnetic and electronic properties of the electron-doped cuprates have been extensively investigated[3]. Compared to the hole-doped side, the long-range AF order persists over a broader doping range on the electron-doped side[3, 36]. Furthermore, the AF wave vector remains commensurate at $(\pi, \pi)$ with doping[1,2], in contrast to stripe-ordered hole-doped materials[25].

Previous electronic structure studies using angle-resolved photoemission spectroscopy (ARPES) have identified an incipient spectral reconstruction in the form of a $(\pi, \pi)$ folding[37], even when the sample has no long-range AF order near optimal doping[38]. In addition, the material exhibits an antiferromagnetic pseudogap[4,5,6,7,8,9,10], which gradually fills as a function of doping in the range where superconductivity appears[39]. In the superconducting state, previous studies used the leading-edge shift of the ARPES spectra to infer the superconducting gap $\Delta_{SC}$[12,40,41,42,43], since the coherence peak was not observed. There are conflicting results on the momentum dependence of $\Delta_{SC}$: An earlier study suggest that $\Delta_{SC}$ exhibits a maximum near the hot spot momenta[12], consistent with a Raman study[11], while other ARPES studies show a simple d-wave form[42,43]. Furthermore, there is a debate as to whether the putative maximum in the leading-edge shift is solely due to superconductivity[44], as the presence of the pseudogap at the hot spots makes the determination of the superconducting gap through the leading-edge shift unreliable. The absence of the superconducting coherence peaks in ARPES measurements thus far has prevented the positive identification of superconductivity-related states. The conflicting results and interpretations in the literature, as well as our desire to solve the pairing dilemma outlined in the abstract, lead us to carry out a comprehensive investigation of this material.

In the following, we study the electronic properties of $Nd_{1.85}Ce_{0.15}CuO_4$ (NCCO) in both the normal and superconducting states. We begin by examining the normal state electronic structure of NCCO, focusing on the presence of the antiferromagnetic pseudogap near the hot spots. Owing to the much-improved experimental conditions and sample preparation processes outlined in methods section II, we were able to observe faint dispersive features inside the AF pseudogap which form the gossamer Fermi surface, and the emergence of Bogoliubov quasiparticles from these states below the superconducting transition temperature ($T_c$). The observation of the superconducting coherence peak allows us to ascertain the intrinsic nature of these in-gap states and to accurately measure the superconducting gap size. The result of these measurements establishes the presence and importance of two sectors of states, providing a route to reconcile the dilemma of maximum superconducting gap occurring where the normal state spectral weight is suppressed the most. Finally, we present a theoretical proposal that explains the experimental observations. By combining these experimental and theoretical approaches, we provide a comprehensive picture of the inner workings of AF spin fluctuations.

**Normal state electronic structure**

We first investigate how short-range antiferromagnetism affects the electronic structure of NCCO in the absence of superconductivity. Fig. 1a-f shows the momentum dependence of the electronic structure of NCCO at x=0.15 in the normal state. Despite the lack of long-range AF order at this doping[38], we observe distinct signatures of electronic structure reconstruction (red dots in cuts 1-4 and red arrow in cut 6, see also extended data Fig. 1) near the AF zone boundary, consistent with previous studies[37]. This reconstruction refers to the ($\pi$, $\pi$) folding of the electronic states. We also observe a dispersion anomaly (green dots) at higher binding energies, which may be attributed to electron-phonon coupling in view of its energy scale[45]. Towards the zone boundary, the two features (red and green dots) become broad and merge into a single broad peak (Fig. 1e and 1h), likely due to increased scattering rates[37]. In the following by "AF pseudogap" feature we refer to the dispersion anomaly (red dots) that bends back the main band. We leave the mechanism responsible for the green dots and its implications for an upcoming study.

Most importantly, we can extract a well-defined dispersion (black traces in Figs. 1a-e obtained from fitting the momentum distribution curves, see also extended data Fig. 3) within the AF pseudogap and observe a clear $E_F$ crossing (orange dots in Fig. 1h) despite the existence of the AF pseudogap. This feature suggests the persistence of a "gossamer" large Fermi surface within the AF pseudogap (red dots). In Fig. 1g we highlight the gossamer Fermi surface in orange. Apparently, it approximates the large Fermi surface without the AF reconstruction. Near the Brillouin zone boundary, we also observe both reconstructed and unreconstructed states, supporting the existence of a gossamer band. Though we note that other factors such as an anisotropic scattering rate may play a contributing role in shaping the spectra[37].

For comparison, we also show the Fermi surfaces of an underdoped 11% sample (Fig. 1i) and an overdoped 19% sample (Fig. 1j), both with an absence of superconductivity. For the underdoped sample with long range AF order, the Fermi surface is dominated by the reconstructed electron pocket centered at (0, $\pi$) with almost no states in the gossamer Fermi surface. For the overdoped sample with no spectral signature of AF, the gossamer Fermi surface has evolved into a full Fermi surface with a Fermi surface volume corresponding to 19% doping.

**Bogoliubov quasiparticles**

After obtaining a clear picture of the electronic structure in the normal state, we proceeded to investigate the superconducting state. Fig. 2a displays the raw data of the ARPES spectra along an arc on the unreconstructed large Fermi surface. It is important to note a low energy spectral feature appearing around ~10 meV below $E_F$ (indicated by the blue arrow). The binding energy of this feature reaches a maximum near the hot spot. To further examine this low energy feature, Figs. 2b and 2c show a cut through the hotspot at temperatures below (7 K) and above (40 K) the $T_c$ of 25 K. A coherence peak (the blue arrow) is clearly visible at 7 K and disappears at 40 K. Figs. 2d and 2e further demonstrate the detailed temperature dependence of the symmetrized EDCs at the hot spot Fermi momentum ($k_F$). The gap associated with the coherence peak closes around 25 K. Fig. 2f shows the extracted low energy gap as a function of temperature, with the temperature-dependent magnetic susceptibility (black curve) also shown. These data unambiguously demonstrate that the coherence peak feature is associated with the Bogoliubov quasiparticle. From Fig. 2a, the largest superconducting gap (~10 meV) is about an order of magnitude smaller than the energy scale associated with the AF pseudogap spectral weight suppression (~ 100 meV) at the hot spot. To clarify the momentum dependence, we note that the AF pseudogap is particle-hole symmetric only at the hot spots. For momenta near the zone diagonal, the Fermi energy is closer to the lower branch of the gapped density of states. This explains why in Fig. 2a the AF pseudogap has an apparent momentum dependent gap edge.

After establishing the superconducting origin of the low energy peak, we investigate its momentum dependence. Fig. 3a shows the EDCs at $k_F$ on the electron pocket and along an arc of gossamer Fermi surface, where we observe Bogoliubov quasiparticle peaks at most momenta (blue dots). However, near the zone diagonal ($\theta \sim 0$), the identification of the presumed superconducting gap node is made difficult by the presence of the AF pseudogap (red dots). The temperature dependence shown in Extended Data Fig. 5 corroborates the AF origin of the zone diagonal gap, which persists significantly above the bulk $T_c$ of 25 K. It is important to note that the gapping of the node by AF has been observed in other electron-doped cuprates[46,47], but since it is derived from the AF pseudogap this "nodal gap" did not invalidate the d-wave pairing symmetry.

The momentum dependence of $\Delta_{SC}$ is determined by fitting the low-energy gap using a phenomenological model (see methods section IV). The extracted gap values are plotted in the top panel of Fig. 3b, along with the corresponding low-energy spectral weight in the bottom panel. Despite the lowest normal state spectral weight at the hot spot, it exhibits the largest $\Delta_{SC}$ of ~11 meV, corresponding to a $2\Delta/k_B T_c \sim 10$ and displays a prominent Bogoliubov quasiparticle peak (Fig. 3c). In contrast, at the zone boundary, where the ($\pi$, $\pi$) folding leaves a reconstructed Fermi surface in the normal state (Fig. 1f), we observe a $\Delta_{SC}$ of ~ 6 meV, corresponding to a $2\Delta/k_B T_c \sim 5$, slightly larger than the weak-coupling BCS value[13]. Furthermore, the Bogoliubov quasiparticle peak at the hot spot is much more evident than that at the zone boundary after normalization (Fig. 3c). These results definitively explain the dilemma discussed earlier – the superconducting gap appears on the gossamer Fermi surface inside the normal state AF pseudogap.

We have also performed specific heat measurements (Extended Data Fig. 6) that reveal a small superconducting transition anomaly compared to the expected BCS value derived from the bare band structure. This is comparable to the small specific heat anomaly in previous measurements at similar doping[48] and consistent with the low spectral weight on the gossamer Fermi surface from which superconductivity arises. Details of the calorimetry measurements are presented in the methods section VIII.

Prior to presenting our theoretical findings, it's reasonable to question whether coexistence of AF reconstruction plus the pseudogap and the gossamer large Fermi surface within the pseudogap can be due to meso-scale separation into an AF-ordered phase and a metallic phase with a large Fermi surface. In the Methods section VI, we present reasons for why this scenario is unlikely. This has motivated us to search for other origins of the in-gap states, discussed in the following.

**Spin orientational fluctuations**

Our approach to understanding the AF pseudogap and the states inside it is based on the concept that the AF order parameter has a modulus and an orientation. The modulus creates the pseudogap and the reconstruction of the electronic structure, while the orientation fluctuation removes the AF long-range order and restores states inside the pseudogap. This proposal is very similar to that proposed in Ref. 49 to account for the states inside the AF pseudogap at the hot spot. This assumption is consistent with previous neutron scattering studies[50,51,52], where the low energy slow spin fluctuations show nearly constant total moment with doping but a gradual diffusion of spectral weight into inelastic channels. Unlike in Ref. 49, the goal of our theoretical model is to show the AF spin orientation fluctuations can restore the gossamer Fermi surface – the large Fermi surface without the AF reconstruction, and to show Cooper pairing can occur on this gossamer Fermi surface.

To test this hypothesis, we carry out numerical simulations under the assumption that the orientation fluctuation occurs at a much slower time scale than the photoemission process. This allows us to calculate the ARPES spectra under different orientation configurations and then average over these configurations (2400 in total). These configurations were generated using the Boltzmann weight of a classical 2D Heisenberg model with different correlation lengths (see details in the methods section).

Fig. 4a and 4b shows the calculated Fermi surfaces at long and short correlation lengths ($\xi_{AF}$) of the spin orientation. At $\xi_{AF}$ much longer than the numerical system size (which is $100 \times 100$ lattice sites), the Fermi surface appears reconstructed, with very little density of states at $E_F$ near the hot spots and zone diagonals. As $\xi_{AF}$ is reduced to ~25 lattice sites, comparable to the measured $\xi_{AF}$ from neutron scattering experiments for this doping and at low temperatures[38], the states along the large Fermi surface are partially restored. We further analyze the energy-momentum spectra across different momenta in Fig. 4c-f. While there is spectral reconstruction (red arrow) at the zone boundary (Fig. 4f), the spectral weight suppression and the residual spectral weight within the AF pseudogap at the hot spot (Fig. 4d cut 2) are evident.

In the above approach, the electrons are assumed to follow the AF orientation fluctuation adiabatically (analogous to the Born-Oppenheimer approximation for phonons). Cooper pairing is triggered by the breaking of this adiabaticity. We assume that after integrating out the fast orientation fluctuations, a short-range AF interaction is generated, which mediates Cooper pairing. Here, in order to avoid a circular argument, Cooper pairing is implemented in the real space by nearest-neighbor d-wave pairing. Fig. 4h shows the calculated superconducting ARPES spectra at the hot spot, where the Bogoliubov quasiparticle coherence peak within the AF pseudogap is evident. Comparing Figs. 4h and 4j with the experimental data in Figs. 4g and 4i, we observe a qualitative agreement. The reader can consult the Methods section V for more details.

**Low energy gossamer states drive coherent superconductivity**

The totality of our results firmly establishes the existence of two sectors of states in superconducting NCCO – states reconstructed and pseudo-gapped by the AF correlations, and states forming the gossamer large Fermi surface within the AF pseudogap. Near the hot spots, these two sectors of states are well-separated in energy. Here, we observe the largest $\Delta_{SC}$ and the most well-defined coherence peaks despite the most prominent spectral weight depletion, suggesting that the states within the gossamer Fermi surface dominate pair formation. The remarkable emergence of coherent Bogoliubov quasiparticles from a phenomenologically incoherent spectra at the hot spot (Fig. 1h) suggests that there is a hidden coherent nature of the gossamer states. On the other hand, the reconstructed states and unreconstructed gossamer states near $E_F$ merge together towards the Brillouin zone boundary. It is an interesting open question how these states interact and lead to a small superconducting gap near the Brillouin zone boundary. One intriguing possibility is that the reconstructed electron pocket centered at $(0, \pi)$ is proximitized by the strong-coupling superconductivity of the gossamer states within the AF pseudogap.

The notion that the gossamer states drive pair formation and phase-coherent superconductivity is consistent with the phase diagram of the electron-doped cuprates[3]. In retrospect, the bulk-superconducting phase seemingly emerges with the appearance of the gossamer Fermi surface. On the underdoped side, the long-ranged AF order depletes almost all the states within the gossamer Fermi surface (Fig. 1i). On the overdoped side, the gossamer Fermi surface evolves into a full unreconstructed large Fermi surface (Fig. 1j). Apparently, the pairing interaction due to spin fluctuations is diminished and the system is not superconducting (see extended data Fig. 8). These observations suggest that the presence of a gossamer Fermi surface and strong spin fluctuations are required for bulk superconductivity in the electron-doped cuprates.

**Implications for strongly correlated superconductors**

Our study of the electron-doped cuprates sheds light on the competition between AF pseudogap and Cooper pairing within a single low energy effective band. Specifically, the spin-1 electron-hole bound states participate in forming the AF order parameter, leaving the rest of the electrons on either the reconstructed Fermi surface or the gossamer large Fermi surface for pairing. A comprehensive understanding of this phenomenon deepens our understanding of spin fluctuations mediated Cooper pairing, which may subsequently be enhanced by additional interactions such as electron-phonon coupling[35].

In our opinion, understanding the important role played by AF fluctuations in the superconductivity in NCCO is a prerequisite for any meaningful discussion of spin fluctuation mediated pairing in the hole-doped cuprates. Specifically, our results here raise the important question of whether a similar gossamer Fermi surface can also exist within the pseudogap of the hole-doped cuprates. If so, the mysterious Fermi arcs[53] and the gossamer Fermi surface near the $(0, \pi)$ may complete the large underlying Fermi surface. More generally, we believe that our study has broader implications and can be applied to deepen our understanding of other electronic-driven superconductors.

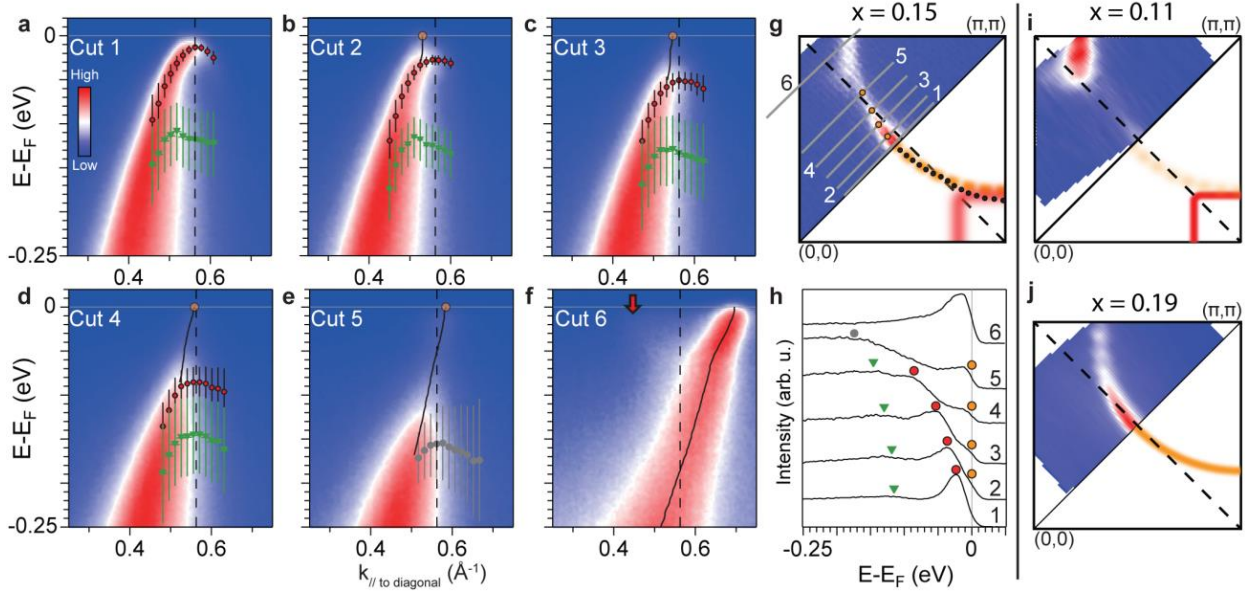

**Fig. 1 | Momentum-dependent normal state electronic structure of $Nd_{1.85}Ce_{0.15}CuO_4$. a-f**, Energy-momentum spectra corresponding to the numbered cuts in **g**. Red dots track the position of the peak associated with the AF pseudogap and green triangles track a higher energy hump feature (see text). Grey dots in **e** track the only observable broad feature in this cut. The black lines in **a-e** tracks the dispersion from fitting the momentum distribution curves (MDCs) at binding energies within the pseudogap in cuts 1-5. Cut 6 passes through the reconstructed electron pocket centered at $(0,\pi)$, and the black line tracks the MDC peaks for the entire energy range shown. Vertical black dashed line indicates the AF zone boundary. The red arrow in **f** highlights the weak $(\pi,\pi)$ folded dispersion branch near the AF zone boundary. **g**, Fermi surface mapping (upper left half) taken by integrating 10 meV of spectra within the Fermi level. The lower right half is a schematic indicating the existence of both the gossamer Fermi surface (orange) and reconstructed Fermi surface (red). Black dashed diagonal line indicates the AF zone boundary. The grey numbered lines indicate the momentum cuts for **a-f**. Black dots overlaid in the bottom right half are the experimentally extracted $k_F$ from the momentum distribution curve (MDC) peaks. **h**, Energy distribution curves (EDCs) at the Fermi momentum ($k_F$) for the cuts 1-6. Black arrow highlights the Fermi surface crossing of the states within the AF pseudogap. Curves are offset vertically for clarity. Orange dots in **a-h** highlight the $k_F$ positions of the gossamer states. **i**, Fermi surface mapping of a 11% doping sample, showing the strong intensity of the Brillouin zone boundary reconstructed electron pocket. **j**, Fermi surface mapping of a 19% doping sample, showing a full large pocket. The extremely overdoped regime is achieved by surface K dosing, and the doping level is estimated from the Fermi surface volume. Data for the 15% sample are taken at MAX IV Bloch beamline. Data for 11% and 19% samples are taken at SSRL beamline 5-4. Measurement temperature for **a-g** is around 25 K, and 7 K for **i** and **j**.

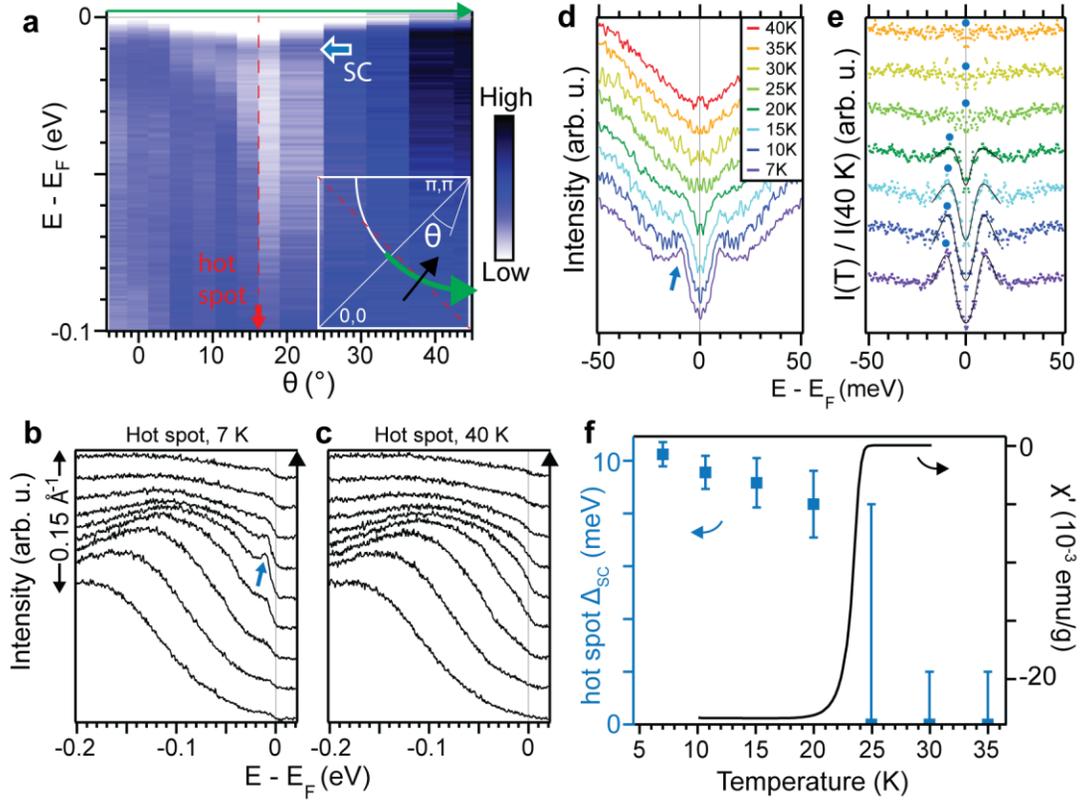

**Fig. 2 | Temperature dependent spectra reveal origin of low energy peak**. **a**, Color-scale plot of the raw EDCs in the superconducting state along an arc on the large Fermi surface indicated by the green arrow in the inset. The blue arrow highlights the low energy peak feature. The red vertical dashed line and arrow indicates the approximate location of the hot spot. Intensity values are normalized by the angular dependent photoemission matrix element extracted from overdoped NCCO spectra, which is minimally affected by AF (see Extended Data Fig. 4). Measurement temperature in **a** is 7 K. **b,c,** EDCs of a cut across the hot spot (indicated by the black arrow in the inset of **a**), at 7 K (**b**) and 40 K (**c**). The cut direction is shown by black arrows beside the panels corresponding to the black arrow in the inset of **a**. The blue arrow highlights the low energy peak feature. **d**, Temperature dependence of the EDCs at the hot spot $k_F$ for different temperatures corresponding to the colored legend in the inset. The blue arrow highlights the low energy peak feature. **e**, Temperature dependence of EDCs normalized by the 40 K spectrum. Black line is the phenomenological fit (see methods section IV) with a constant background. Blue dots highlight the superconducting gap edge. **f**, Left: extracted superconducting gap as a function of temperature. Right: temperature-dependent magnetic susceptibility. Curves in **b-e** are offset vertically for clarity. Data taken at SSRL beamline 5-4.

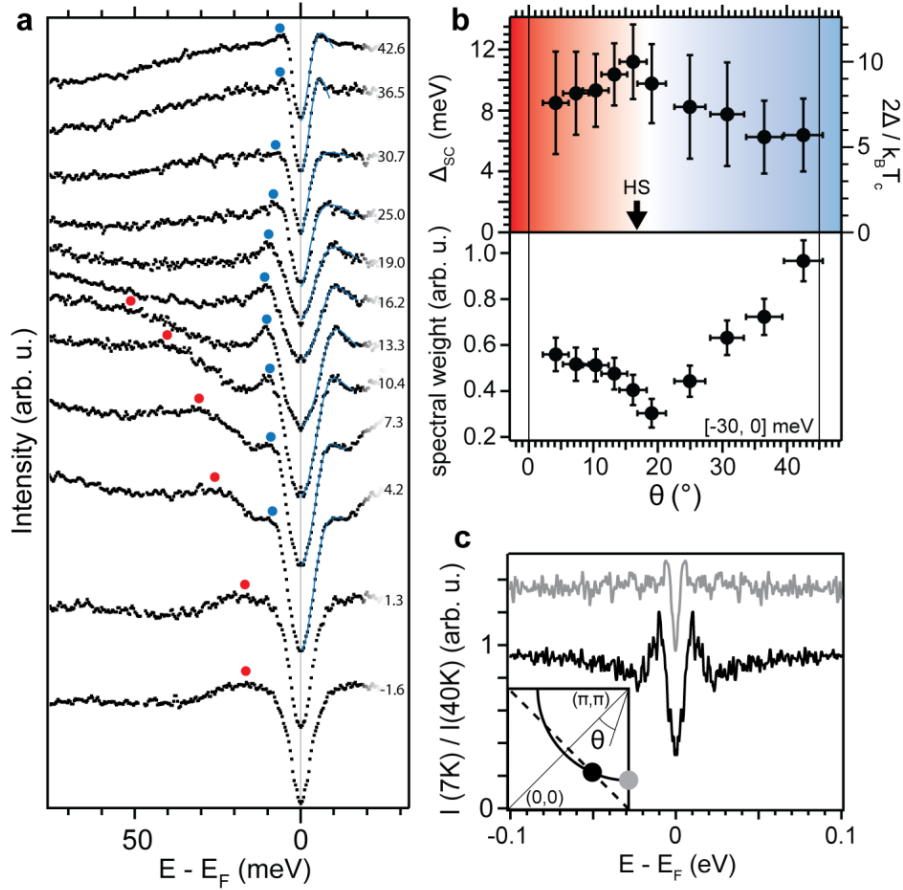

**Fig. 3 | Momentum dependence of Bogoliubov quasiparticles. a,** Momentum dependence of the $k_F$ spectra along an arc on the large Fermi surface. Red dots track the peak feature related to AF pseudogap, and blue dots track the low energy Bogoliubov quasiparticle peak. Numbers to the right indicate angle away from the Brillouin zone diagonal, with the angle θ defined in the inset of **c**. Blue curves overlaying the EDC data in the symmetrized part above $E_F$ are the phenomenological fits of the superconducting gap (see methods section IV). A second order background was used to capture the normal state background spectral weight. Curves near θ = 0 were not fitted as the large size of the AF pseudogap overshadows any small superconducting gap that may be present within. **b,** (top) Momentum dependence of $\Delta_{SC}$ and corresponding $2\Delta_{SC}/k_B T_c$ ratios measured at 7 K. The black vertical arrow indicates the approximate location of the hot spot. Gap data near the Brillouin zone diagonal is not available due to the dominance of the AF pseudogap. Red background highlights the dominance of AF on the low energy spectra near θ = 0, and the blue background highlight the dominance of superconductivity near θ = 45. (bottom) Integrated spectral weight in the range [-30, 0] meV. Intensity values are normalized by the angular dependent photoemission matrix element extracted from overdoped NCCO spectra. **c,** EDCs at the hot spot (black) and Brillouin zone boundary (grey) at 7 K for the same sample, normalized by the respective spectra at 40 K. Brillouin zone location of EDCs shown in the inset. The Brillouin zone boundary curve is offset vertically by 0.2 for clarity.

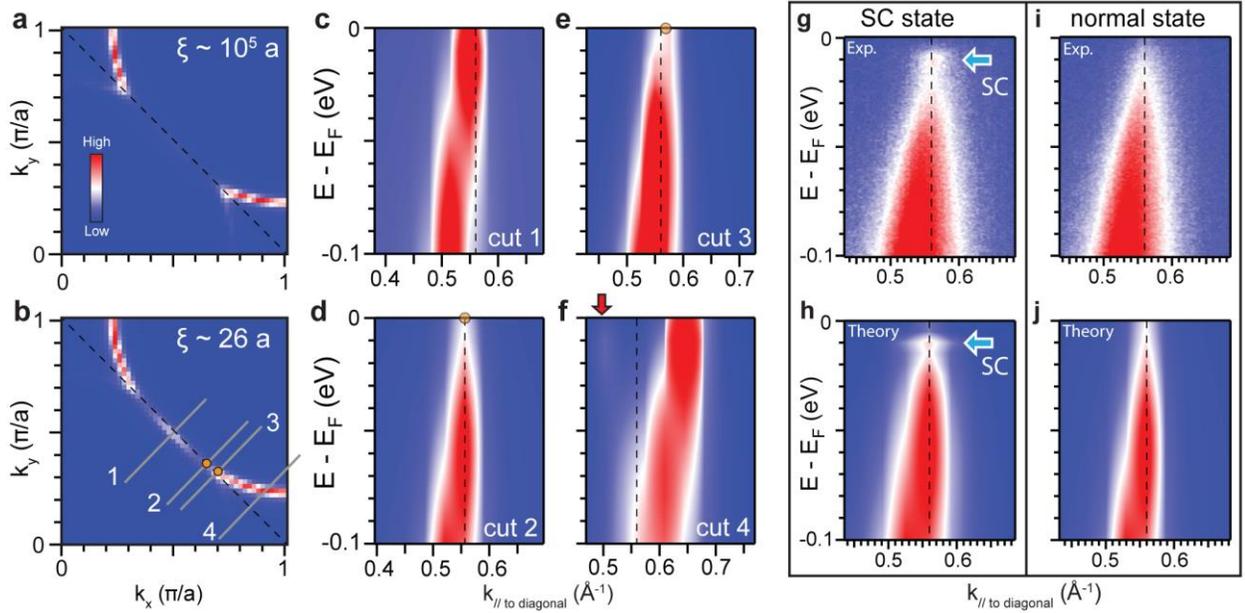

**Fig. 4 | Spin orientational fluctuations restore gossamer states. a,b,** Calculated Fermi surfaces at AF correlation lengths $\xi_{AF}$ of $10^5$ a (**a**) and ~25 a (**b**), where a is the lattice constant. **c-f,** Calculated spectra of cuts indicated by the grey lines in **b**. The red arrow above cut 4 highlights the reconstructed feature near the AF zone boundary. Orange dots highlight the presence of the gossamer Fermi surface within the AF pseudogap. **g,h,** Experimental (**g**) and calculated (**h**) spectra in the superconducting state. The calculated superconducting spectra is obtained by real-space cooper pairing. **i,j,** Experimental (**i**) and calculated (**j**) spectra in the normal state. Blue arrows in **g** and **h** highlight coherence peak of the Bogoliubov quasiparticle. Black dashed lines in all panels indicate the AF zone boundary.

**Methods**

I. Sample synthesis and annealing

$Nd_{2-x}Ce_xCuO_4$ (NCCO) single crystals were grown with the traveling solvent floating zone method with CuO flux in the molten zone. The crystals were annealed at 900 C under flowing Ar gas were optimized to obtain the highest and the sharpest superconducting transition: $T_c \sim 25$ K and $\Delta T_c \sim 2$ K.. The $T_c$ of the samples were characterized in a Physical Property Measurement System using the AC susceptibility method. We note that here the onset of diamagnetism is taken as $T_c$, as that is usually the temperature where the resistivity reaches 0. The width of the transition $\Delta T_c$ is defined here as the temperature range over which the magnetization drops from 90% to 10%. The Ce content of the crystals was characterized by wavelength dispersive spectroscopy in an electron probe microanalyzer.

II. ARPES measurements

ARPES measurements were performed at the Bloch beamline (MAX IV), and beamline 5-4 (Stanford Synchrotron Radiation Lightsource, or SSRL). Preliminary measurements were also carried out at beamline 5-2 (SSRL). Single crystal NCCO samples were mounted on top of copper posts with H20E silver epoxy and Torrseal. Laue back scattering was performed to align the sample in the basal plane. A ceramic top post was mounted with Torrseal for in situ cleaving. 16.5 eV photons were used to obtain an adequate zone boundary intensity while keeping a low photon energy for good energy resolutions. The beam spot size is estimated to be around 30 μm by 18 μm at MAX IV Bloch beamline and about 100 μm by 150 μm at SSRL beamline 5-4. The Fermi level $E_F$ is measured on a reference polycrystal gold and carefully checked for extrinsic sample charging and space charging effects for the superconducting gap measurements. During the superconducting gap measurements at SSRL beamline 5-4, $E_F$ is generally measured about every 30-45 minutes to account for a slow drift of the photon energy on the scale of < 0.5 meV/hour due to monochromator warm up and diurnal variations. We note that once the beamline reaches a stable state, the drift is usually about 0.1 meV/hour or less. The experimental resolution for the superconducting gap measurements is about 4 meV and is determined from fitting the reference gold Fermi cutoff.

We find that a number of factors work in concert to enable the measurement of high-quality spectra and the observation of the Bogoliubov quasiparticle in the electron-dope cuprates:

1. We have synthesized and annealed high quality $Nd_{1.85}Ce_{0.15}CuO_4$ crystals that show $T_c$ (defined by the onset of magnetization drop, usually comparable to when the resistivity reaches 0) of 25-26 K and transition width of ~2 K. A high $T_c$ and narrow transition width, indicative of high crystal quality and uniformity, are required for observing the Bogoliubov quasiparticles.
2. The measurement photon energy selection is important for achieving high counts without compromising the measurement resolution (through effects such as space charging). A high flux, tunable, and highly monochromatic light source, such as Beamline 5-4 at SSRL or Bloch beamline at MAX IV, is required for this measurement.
3. For the superconducting gap measurements, high energy resolution and a highly stable light source is required. We find that the typical photon energy drift (>1-2 meV/h) at most beamlines is not suitable for the superconducting gap measurements on the electron-doped cuprate superconductors, due to the relatively low counts of the valence electrons in the cuprates and therefore long data acquisition times required.

4. During the temperature-dependent measurements, excellent control of the vacuum at different temperatures is required. For Beamline 5-4 at SSRL, we have implemented a local heater that controls the sample temperature without disturbing the temperature of the entire manipulator arm. This way, outgassing during temperature cycling is minimized and the measurement chamber vacuum is kept below $2.5 \times 10^{-11}$ Torr at all times.
5. For improved signal-to-noise ratio, the metal mesh in front of the multi-channel plate detector at Beamline 5-4 of SSRL is removed. Due to the excellent vacuum conditions and µ-metal shielding on the chamber, the mesh that is used to suppress ion feedback noise and shield magnetic fields has minimal benefits.
6. Unlike the commonly-studied hole-doped cuprate $Bi_2Sr_2CaCu_2O_8$ that are flaky with a Van Der Waal BiO-BiO interface, the T′ electron-doped cuprates are rock-like and difficult to cleave. In order to obtain flat cleavage surfaces that show step-like terraces, special cleaving geometries are required. Here, the sample is first shaped into an elongated rectangle, with the long direction along the c-axis. Then Torrseal is applied with the top ceramic post such that the epoxy covers significant portions of the sides of the crystal. This way, the cleaving process applied a uniform fracture force along a small cross-section and is more likely to generate a high quality cleave.
7. Even with the above cleaving method, there may be areas of the cleaved surface that have rugged terrain which does not produce high quality ARPES spectra. A small beam spot combined with spatial photoemission scanning capabilities are required to seek out high quality cleaved spots. For the gap measurements at 5-4 with a slightly larger beam spot, multiple cleaves are usually needed to find a cleaved surface with a larger area of high quality surface.

III. ARPES data processing

To properly process the raw data to the presented data in the figures, several careful calibrations and conversions are required. Standard ARPES data processing procedures[54,55] were performed, including: detector channel inhomogeneity calibration, detector nonlinearity calibration, analyzer slit inhomogeneity and lensing calibration, $E_F$ calibration with respect to an electrically connected polycrystalline gold reference, and emission-angle-to-momentum conversion. Normalizations of ARPES spectral intensities are performed using the high-order light background intensity above $E_F$, well above any significant thermal tail of the Fermi-Dirac distribution. To remove the incoherent scattering background, a reference EDC background far away from dispersive features is subtracted from the spectra.

IV. Phenomenological fitting of the superconducting gap

The superconducting gap is fitted using the phenomenological model in Ref. 56, where the self-energy in the superconducting state has the following form

$$\Sigma(\mathbf{k}, \omega) = -i\Gamma_1 + \frac{\Delta^2}{\omega + \epsilon(\mathbf{k}) + i\Gamma_0}$$

Here, $\Gamma_1$ is the single particle scattering rate, $\Delta$ is the gap magnitude. We note that due to the large energy scale of the AF gap compared to the superconducting gap, the inverse pair lifetime $i\Gamma_0$, used to account for the pseudogap on the hole doped side, can be taken as 0 below $T_c$ when modeling the low energy superconducting peak in NCCO. For the temperature-dependent EDC fittings, the pair-breaking term is left as a free parameter.

V. Spin orientation fluctuation calculations

The AF order parameter is a three-component vector. It has a modulus $\rho_{AF}$ and an orientation $\hat{n}$, namely, $\vec{N} = \rho_{AF}\hat{n}$. This is analogous to the superconducting (SC) order parameter has a modulus and a phase, namely, $\rho_{SC}e^{i\theta}$. In the latter case, the SC gap-opening temperature marks the onset of non-zero $\rho_{SC}$, while the phase coherence temperature marks the onset of $\langle e^{i\theta}\rangle \neq 0$ (and the vanishing of the resistivity). Similarly, in an antiferromagnet, the onset of non-zero $\rho_{AF}$ signals the opening of the AF gap. However, because the AF order occurs at non-zero momentum, non-zero $\rho_{AF}$ generically only partially gap the Fermi surface (unless there is nesting). However even $\rho_{AF}$ is non-zero, whether there is AF long-range order depends on whether $\hat{n}$ is ordered.

In view of the experimental data, we come to realize that in an AF system with non-zero $\rho_{AF}$ but with disordered $\hat{n}$, the Fermi surface can be restored due to the orientation fluctuations. Our theoretical treatment assumes that the orientation fluctuations are slow in comparison with the photoemission time scale, hence we view the measured photoemission spectra as taking snapshots of the electronic structure under different orientation configurations, then average over these configurations. This is analogous to the "Born-Oppenheimer" approximation used to treat phonons. In addition, we treat the breakdown of this approximation as causing Cooper pairing just as in the electron-phonon interaction problem.

Our calculation is done with a tight-binding model that fits the high temperature photoemission data. The tight binding Hamiltonian is given by

$$H_0 = \sum_{i,j,\sigma}(-t_{ij}c_{i\sigma}^+ c_{j\sigma} + h.c) - \mu \sum_{i\sigma} c_{i\sigma}^+ c_{i\sigma}$$

where for the nearest-neighbor/second nearest-neighbor hopping $t_{ij}$ =0.326/-0.0766 eV, respectively; the chemical potential $\mu$ is taken to be -0.04 eV. We then set $\rho_{AF}$ by demanding the zero-temperature AF gap at the hotspot be approximately equal to that measured in ARPES ($\approx 0.1$ eV). Subsequently we generate a set of orientation ($\hat{n}$) configurations, using a 2D classical Heisenberg model as the Boltzmann weight. The Hamiltonian in the presence of the fluctuating AF order parameters is given by

$$H = H_0 + \rho_{AF} \sum_{i,\alpha,\beta}(-1)^i c_{i\alpha}^+ (\hat{n}_i \cdot \vec{\sigma}_{\alpha\beta}) c_{i\beta}.$$

Here $(-1)^i$ is the staggered factor associated with AF, $\vec{\sigma}$ are the Pauli matrices and $\alpha, \beta = x, y, z$. We characterize each set of configurations by its orientation correlation length $\xi_{AF}$ (which is controlled by the coupling strength that enters the Heisenberg model). Since we need to compute the electron spectral function under an arbitrary orientation configuration, such calculation needs to be carried out numerically on a $L \times L$ lattice ($L_{max} = 100$ for Fermi surface mapping, and 60 in the energy-momentum cuts). We broaden the discrete energy levels at finite $L$ by an energy-dependent factor $\Gamma(E) = 0.002 \text{ eV} + 0.4 \times E$. The results associated with different configurations are then averaged over 2400 orientation configurations. To study the superconducting pairing, we assume after integrating out the fast orientation fluctuation a short-range AF interaction is generated which causes Cooper pairing. Cooper pairing is implemented in the real space by a nearest-neighbor d-wave pairing with pairing amplitude $\Delta_{ij} = 0.01\text{eV}$.

VI. Phase separation is an unlikely origin of in-gap states.

Doping and magnetic phase separation could produce regions of AF order and superconductivity. The resulting ARPES spectra may present as a combination of reconstructed and unreconstructed Fermi surface. In the following we present arguments disfavoring the coexistence of metallic spatial regions and AF spatial regions with different doping densities being the primary factor governing the behavior observed.

1. Analysis of the momentum distribution curves near the hot spot shows that the dispersions of the high and low energy states are consistent with the same doping (see Extended Data Fig. 7) thereby excluding significant doping inhomogeneity.
2. Neutron scattering results[57] of NCCO at the same doping revealed a spin gap at low energies and temperatures. Moreover, there is no evidence of the spin excitations below the spin gap within the experimental resolution. This indicates that there are no significant spatial regions where there is well-defined AF order.
3. Field-dependent neutron scattering results[58] show that the dynamical magnetic correlations are long-ranged, spanning vortex cores and superconducting regions. This implies that the magnetic fluctuations are uniform at the scale of the superconducting coherence length.
4. Cu NMR results[59] show a lack of a low temperature "wipe out" effect in the closely related compound $Pr_{1.85}Ce_{0.15}CuO_4$, indicating that there is no slowing down of the spins at low temperatures. Since the NMR signal is sensitive to the local magnetic field, this result suggests that as far as magnetism is concerned the sample appear homogeneous.
5. In the scenario of mesoscopic phase separation, it is the volume fraction that changes as a function of doping, rather than the chemical potential. However, the chemical potential shifts continuously with doping in NCCO[60], indicating that phase separation is not significant.

VII. Specific heat measurements

Specific heat measurements were carried out in a differential membrane-based nanocalorimeter[61] at Stockholm University. A small piece of the sample was broken off from the larger crystals used for the ARPES studies and mounted with Apiezon grease. The background of the empty cell and grease were calibrated. A magnetic field was used to push the specific heat jump from ~25 K to a lower temperature, such that the zero-magnetic-field jump magnitude can be detected from the difference between the zero field and finite field curves (Extended Data Fig. 6). The absolute value of the molar specific heat was estimated from the magnitude of the low temperature Nd moment Schottky anomaly signal.

VIII. Expected specific heat jump from band structure

The electronic specific heat is related to the density of states through

$$C_e = \frac{\partial S_e}{\partial T}$$

Where $S_e$ is the quasiparticle entropy

$$S_e = k_B \int [-f \ln f - (1-f) \ln(1-f)] D(E) dE$$

Where $k_B$ is the Boltzmann constant, $f$ is the Fermi-Dirac function, and $D(E)$ is the quasiparticle density of state. $D(E)$ can be calculated from the tight binding band parameter by integrating the states near the Fermi level in momentum space

$$D(E) = \frac{1}{4\pi^3} \int_{FS} \frac{dk}{\partial E(k)}$$

The results for the Sommerfeld coefficient $\gamma$, or C/T, for $Nd_{2-x}Ce_xCuO_4$ and $La_{2-x}Sr_xCuO_4$ are shown in Extended Data Fig. 6, using the tight binding parameters for NCCO and LSCO respectively. Here, as the presence of the AF reconstruction affects the measured dispersion in underdoped and optimally doped NCCO, the bare band tight binding parameters is approximated by the overdoped side where the effect of AF is minimal. The tight binding parameters used for the DOS calculations here are $\mu = -0.065$, $t = 0.2497$, $t' = -0.1036$, extracted from the heavily surface-dosed spectra (similar to that in Extended Data Fig. 4). Using the parameters from a different doping is justified for the bare band expectation as the density of states does not vary significantly in this doping regime, such that a slight shift in the doping and therefore chemical potential level affects the electronic $\gamma$ minimally.

This result does not take into consideration of the AF pseudogap and reconstruction, which may partially gap out states and affect the actual density of states. By comparing the expected specific anomaly $\Delta C/T$ of the bare band to the experimental value, one can estimate the effects of the AF pseudogap. At x=0.15, the calculated $\gamma$ is about 2.6 mJ mJ/molK$^2$ (Fig. S8). Combined with the BCS expectation for a d-wave superconductor[62,63]

$$\Delta C_{BCS}/T = 1.43\gamma$$

The tight binding band specific anomaly is estimated to be about 3.7 mJ/molK$^2$. Compared to the measured experimental $\Delta C/T$ of 1.2 mJ/molK$^2$ (Extended Data Fig. 6), the calculated $\Delta C_{BCS}/T$ is more than twice as large, indicating that a significant portion of the spectral weight is missing in the superconducting transition. We have also calculated $\gamma$ using the x=0.15 fitted tight binding parameters ($\mu = -0.04$, $t = 0.326$, $t' = -0.0766$), which includes renormalization effects from AF but not the spectral weight suppression. This method produces about 3.1 mJ/molK$^2$ at 15% doping, still significantly larger than the experimentally measured value of 1.2 mJ/molK$^2$.

**Data availability statement**

The data presented in this work is available at the Stanford Digital Depository (Link to be added).

## Acknowledgements

We thank S. A. Kivelson, L. Taillefer, Y. Wang, B. Moritz, and Y. Zhong for fruitful discussions. Technical assistance and valuable discussions with C. Polly, B. Thiagarajan and H. Fedderwitz, are gratefully acknowledged. We acknowledge SSRL and MAX IV for beamtime on Beamline 5 (SSRL) under the approved program and Bloch Beamline (MAX IV) under Proposal 20210256, respectively. The work at Stanford and SSRL was supported by the US Department of Energy, Office of Science, Office of Basic Energy Sciences, Materials Sciences and Engineering Division, under Contract DE-AC02-76SF00515. D.-H.Lee. was supported by the US Department of Energy, Office of Science, Basic Energy Sciences, Materials Sciences and Engineering Division, contract no. DE-AC02-05-CH11231 within the Quantum Materials Program (KC2202). The work at KTH and MAX IV was supported by Swedish Research Council grants 2019-00701 and 2019-03486, as well as The Knut and Alice Wallenberg foundation grant 2018.0104.

## Author contributions

K.-J.X., M.H., O.T., and Z.-X.S. conceived the experiment. K.-J.X., J.H. C.R.R., and Y.S.L synthesized the samples. K.-J.X. M.H., and D.-H.Lu performed the ARPES measurements at SSRL. Q.G., C.L., and M.H.B performed the ARPES measurements at MAX IV. A. R. performed the specific heat measurements. K.-J.X., Q.G., M.H., S.-D.C., Y.H., T.P.D., D.-H.Lee, O.T., and Z.-X.S. analyzed and interpreted the ARPES data. Z.L. and D.-H.Lee conceived the theoretical model and performed numerical calculations. K.-J.X., M.H., D.-H.Lee and Z.-X.S. wrote the manuscript with input from all authors.

## Competing interest declaration

The authors declare no competing interests.

## Additional information

## Supplementary information

**Correspondence and requests for materials** should be addressed to Dung-Hai Lee, Oscar Tjernberg, or Zhi-Xun Shen

**Extended data**

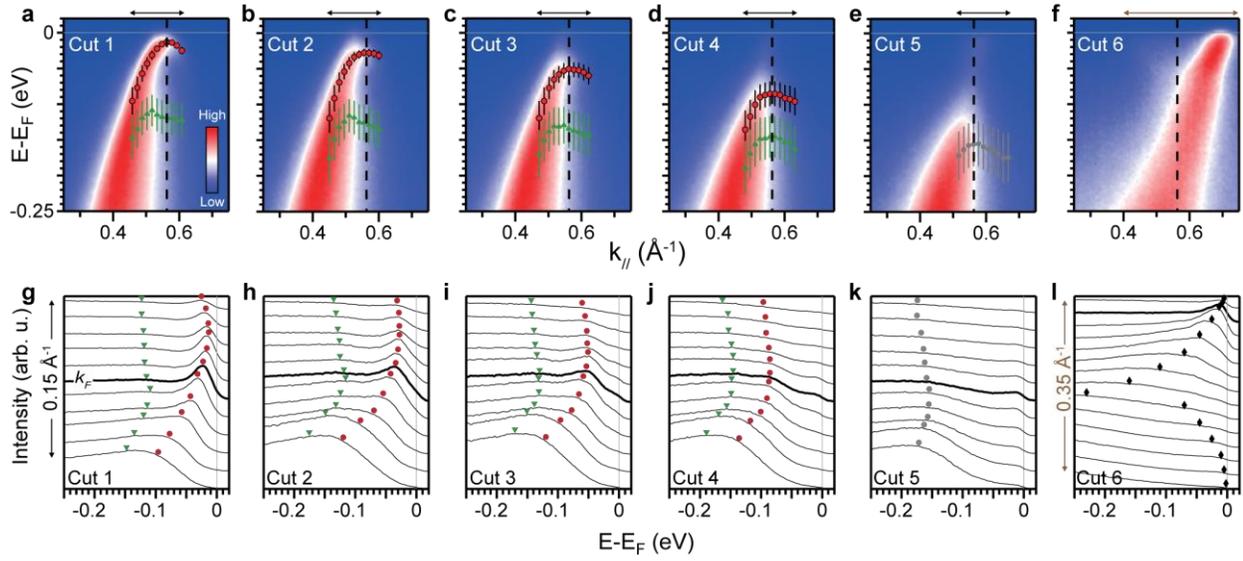

**Extended Data Fig. 1 | Energy distribution curves of the normal state spectra in Nd$_{1.85}$Ce$_{0.15}$CuO$_4$. a-f,** energy-momentum spectra from the corresponding cuts 1-6 in main Fig. 1g. Red dots track the position of the peak associated with AF and green triangles track a higher energy hump feature (see main text). Grey dots in (**e**) track the only observable broad feature in this cut. Vertical black dashed line indicates the AF zone boundary. Grey line indicates the Fermi energy $E_F$. **g-l,** energy distribution curves (EDCs) of the momentum region indicated by the black double arrows in the respective cuts in **a-f**. Red, green, and grey dots indicate the same respective features as **a-f**. Black diamonds in **l** indicates the zone boundary dispersion and reconstructed electron pocket. The thick black line indicates the Fermi momentum $k_F$, and the grey vertical line indicates $E_F$.

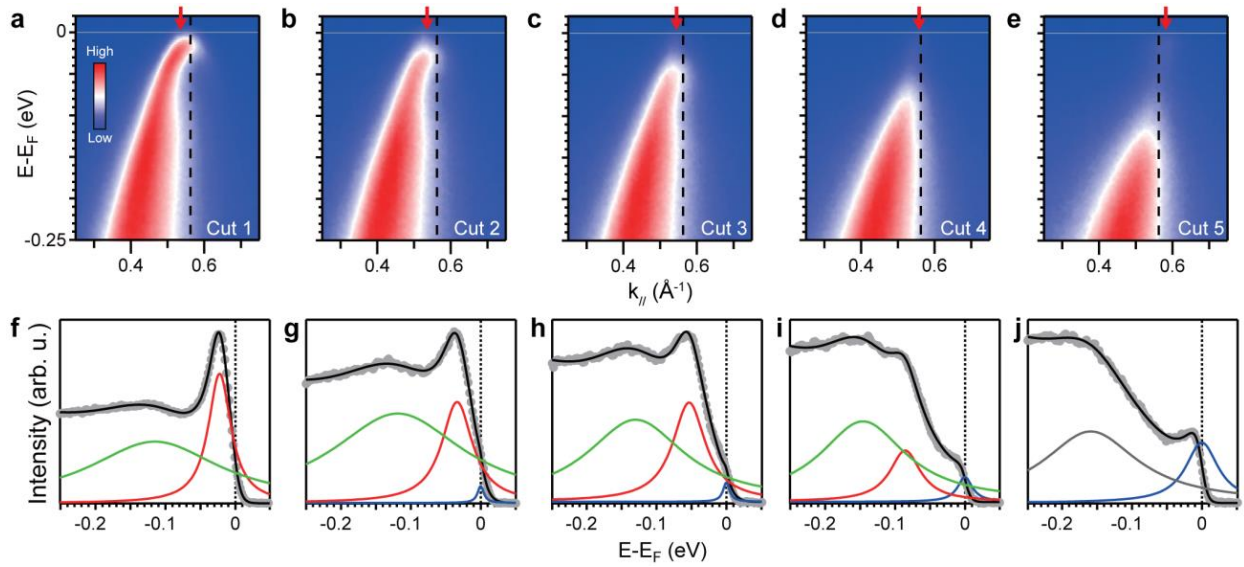

**Extended Data Fig. 2 | Spectral fittings in the normal state spectra. a-e**, Normal spectra for cuts 1-5 that are the same as that in main Fig. 1g. Small red arrow points to the Fermi momentum. **f-j**, Fits of the EDC curves. Grey dots are the raw $k_F$ EDC data, with the fittings shown by the overlaying black curve. The EDCs are fitted with 3 Lorentzian in **g-i** and 2 Lorentzian for **f** and **j**. Each fit is also convolved with the Fermi-Dirac function.

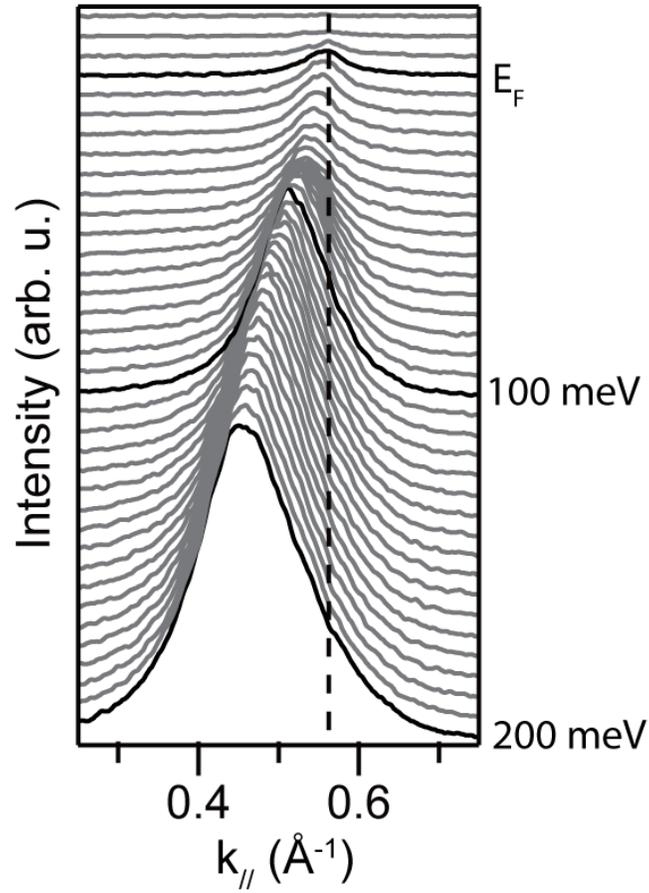

**Extended Data Fig. 3 | Momentum distribution curves of gossamer states**. MDCs at the hot spot, same as cut 4 in the main text Fig. 1, showing peaks within the AF pseudogap (which has a broad gap edge at around 100 meV). The dispersion of the gossamer states is extracted by fitting the MDC peaks with Lorentzian peaks.

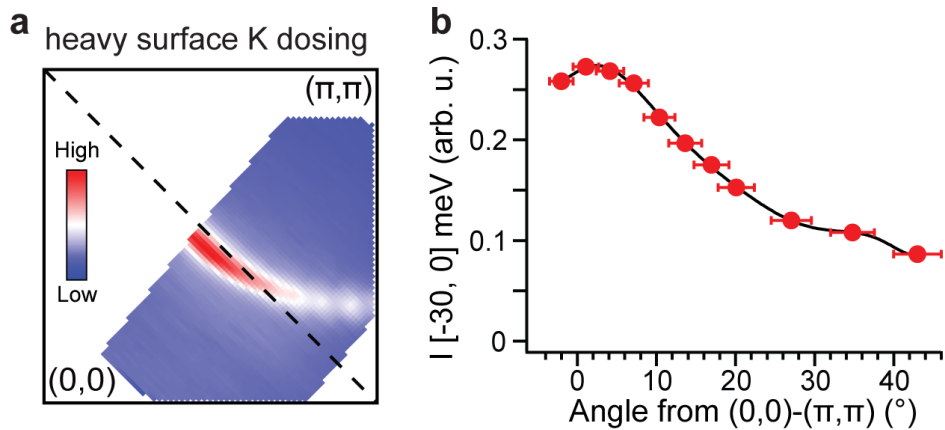

**Extended Data Fig. 4 | Momentum-dependent matrix element normalization. a**, Fermi surface mapping of an NCCO sample where the surface was heavily dosed with K atoms, such that there are no visible AF spectral signatures. **b**, angular dependence of the $k_F$ spectral weight of the surface-K-dosed sample (red dots) as in **a**. The black curve is a high order polynomial fit to extract the angle-dependent normalization factor.

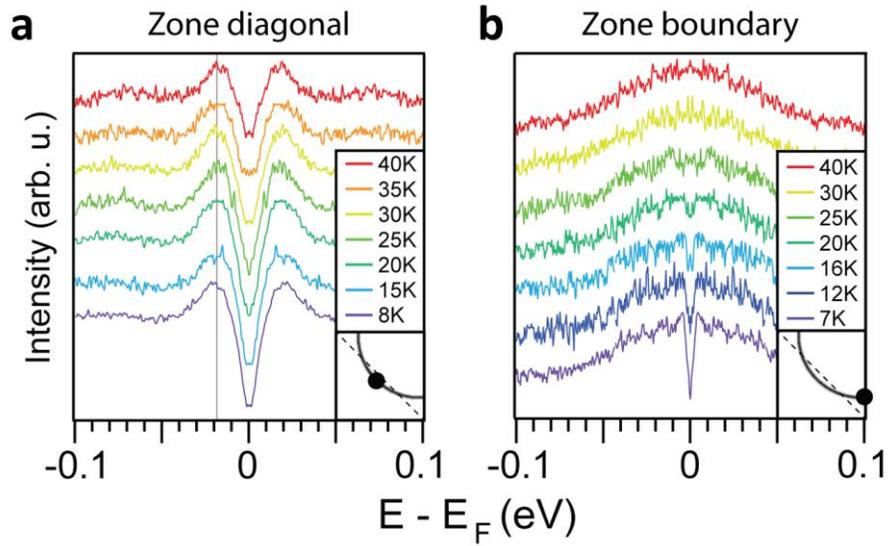

**Extended Data Fig. 5 | Temperature dependence of Brillouin zone diagonal and Brillouin zone boundary spectra for $Nd_{1.85}Ce_{0.15}CuO_4$. a,b,** Temperature dependence of the $k_F$ EDCs at the Brillouin zone diagonal (**a**) and Brillouin zone boundary (**b**), with Brillouin zone position shown in the respective insets. Measurement temperatures are indicated by the colored legends. The grey vertical line in (**a**) is a guide to the eye highlighting the persistence of the AF feature through the superconducting transition temperature of 25 K.

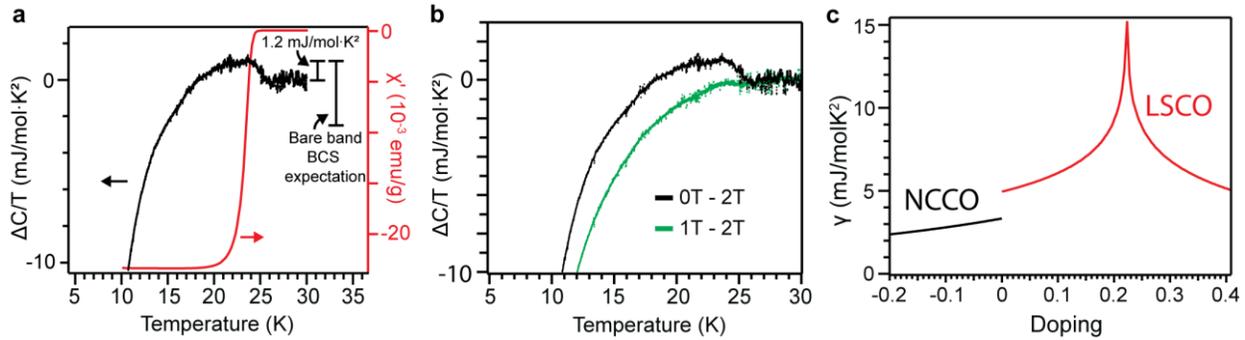

**Extended Data Fig. 6 | Specific heat reveals small superconducting transition anomaly**. **a,** Difference specific heat (left) and magnetic susceptibility (right) as a function of temperature. The difference specific heat is extracted by subtracting the zero field curve by the 2T curve. The low temperature drop off is from the field dependence of the Nd Schottky anomaly. The bare band expectation based on Bardeen-Cooper-Schrieffer (BCS) theory is derived from the calculated electronic specific heat in **c**. **b,** comparison of the zero field and finite field C/T curves for extraction of the electronic specific heat. 2T is enough to suppress the transition to a lower temperature such that the full height of the jump is preserved in the difference curve. **c,** Calculated electronic specific heat from the bare band tight binding parameters, for NCCO (black curve) and LSCO (red curve). See methods sections VII and VIII for additional information of the electronic specific heat calculation and derivation of the expected BCS specific heat jump size.

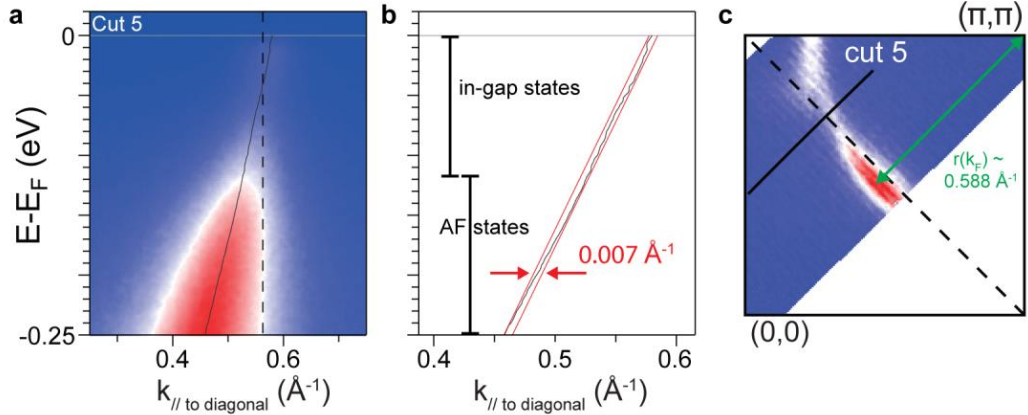

**Extended Data Fig. 7 | Excluding significant doping inhomogeneity. a**, ARPES spectra of the cut corresponding to the cut shown in **c**. Diagonal black line is the dispersion extracted from fitting the momentum distribution curves (MDCs). **b**, Dispersion extracted from MDC fitting. The red lines are bounds on the momentum deviation of the dispersion. **c**, Fermi surface mapping constructed using intensity from ±10 meV of $E_F$. Green line indicates the approximate radius of the large Fermi pocket. We can place an upper bound on the doping uncertainty by the momentum bounds in **b**. Here, the effect of the AF reconstruction is broad enough to facilitate the extraction of an approximately linear dispersion in this energy range. The intensity between $E_F$ and $E_B \sim 100$ meV is from the in-gap residual states, and the intensity at $E_B > \sim 100$ meV is dominated by the AF states. If the AF and in-gap states originate from distinct doping regions, then we expect the dispersions of the in-gap states and AF states to be offset in momentum. The deviation from a linear dispersion is bounded to be about $\Delta k \sim 0.007$ Å$^{-1}$. We note that while the individual MDC widths are larger than this momentum uncertainty, the collective noise in the dispersion fittings show an error smaller than $\Delta k \sim 0.007$ Å$^{-1}$. This momentum uncertainty can be translated to a Fermi surface volume uncertainty by the following approximation $\Delta A_{FS} \sim 2\pi r(k_F) \cdot \Delta k = 0.025 \pm 0.003$ Å$^{-2}$, where $r(k_F)$ is the approximate radius of the large Fermi surface that is nearly circular in this doping regime. We note that the uncertainty in $r(k_F)$, which we use a generous estimate of ±0.05 Å$^{-1}$, is about 10% and linearly affects $\Delta A_{FS}$. The doping uncertainty is the $\Delta n = \Delta A_{FS}/A_{FS} = 0.01 \pm 0.001$ Å$^{-2}$. The doping inhomogeneity upper bound is about 1% (or ±0.5% from the nominal 15% doping), and thus excludes significant doping phase separation in our measured samples.

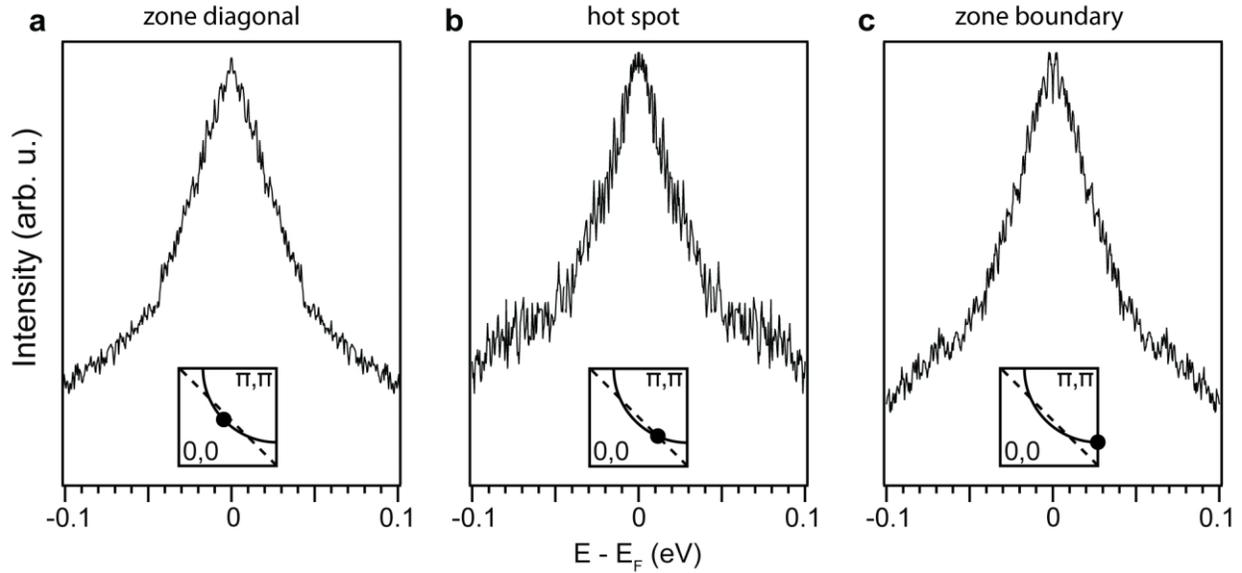

**Extended Data Fig. 8 | Symmetrized spectra in overdoped sample reached by surface K dosing**. **a-c**, symmetrized EDC at the zone diagonal (**a**), hot spot (**b**), and zone boundary (**c**) $k_F$. The momentum location of the EDC is shown in the respective insets. Here, little to no signature of the incipient antiferromagnetic gap is observed anywhere in momentum. No superconducting gap is observed within the experimental temperature and resolution. We note that while the experimental resolution is ~4 meV, this number is defined by the broadening of the Fermi Dirac cutoff, and typically one can observe gap features much smaller than the experimental resolution. Here, the extremely overdoped side is reached via surface K dosing, which introduces additional electrons into the system. From the Fermi surface volume, we estimate the doping to be about 19%.